\documentstyle[11pt]{article}

\def\lapproxeq{\lower .7ex\hbox{$\;\stackrel{\textstyle <}{\sim}\;$}}
\def\gapproxeq{\lower .7ex\hbox{$\;\stackrel{\textstyle >}{\sim}\;$}}

\begin{document}

\titlepage
\begin{flushright} MC-TH-95/23 \\ ANL-HEP-PR-95-89\\ 
DESY-95-253\\
December 1995
\end{flushright}

\begin{center}
\vspace*{1cm}
{\large{\bf  Diffractive Production of Vector Mesons at Large $t$}}
\end{center}
\vspace*{.5cm}
\begin{center}
J.Bartels$^1$, J.R. Forshaw$^2$, H.Lotter$^1$ and
M.W\"usthoff$^3$\footnote{Work partly supported by the U.S.Department of
Energy, Division of High Energy Physics, Contract W-31-109-ENG-38}
\\
$^1$Institut f\"ur Theoretische Physik, \\ Universit\"at Hamburg, \\ Hamburg,
Germany. \\
$^2$ University of Manchester, \\ Manchester, M13 9PL, England. \\
$^3$ High Energy Physics Division, \\Argonne National Laboratory, USA.
\end{center}

\vspace*{0.5cm}

\begin{abstract}
The cross section for elastic vector meson production in photon-proton
interactions at large $t$ is considered using the exact
analytic solution of the BFKL equation in the azimuthally symmetric $n=0$ limit.
We use a non-relativistic model for the vector meson production and find
a small shrinkage in the $t$-distribution with increasing energy. 
\end{abstract}

\newpage

\section{Introduction}
We study the process $\gamma p \to V+X$ (where $V$ is a vector meson and
$X$ denotes the products of the proton dissociation) within
the Regge limit assuming a large moment transfer, $ s \gg -t \gg Q_0^2$,
 i.e. large angle meson production at high centre-of-mass energies.
$Q_0$ denotes the typical hadronic scale of $\sim 1$ GeV. 
The largeness of $-t$ allows us to apply the solutions to the nonforward 
BFKL-equation which are given as a set of conformally invariant 
eigenfunctions \cite{lip} (BFKL-Pomeron). 
This process has been studied previously in \cite{fr}, where 
an analytic solution of the BFKL equation was given for asymptotically large
$s$ and for $-t$ either very large or very small (relative to the larger
of the meson mass and photon virtuality). Comparisons
with an iterative numerical solution were also made. An appreciable cross 
section was found and the prospect of revealing the BFKL dynamics at HERA 
was established.

Here we derive a formula for the cross section which is based on
the solution of the nonforward BFKL equation in the azimuthally
symmetric approximation. We present a detailed numerical
evaluation of this formula.

Throughout we work in the approximation that the light-cone wavefunction 
of the vector meson is a $\delta$-function, which partitions the 
light-cone momenta and the transverse momenta of the 
quark-antiquark pair (from the photon) equally. This should be appropriate
for heavy vector mesons (e.g. $J/\Psi$ and $\Upsilon$) and possibly
also for the lighter mesons ($\rho$, $\omega$, $\phi$), however,
the choice of the wave function is not uniquely determined. One could
also think of a softer distribution than a $\delta$-function. 

For the coupling of the Pomeron to the proton we use 
the Mueller-Tang prescription \cite{mt} where the 
pomeron couples to a single parton only.
This is justified \cite{bart} since, in the case considered 
here, the size of the Pomeron ($1/\sqrt{-t}$) 
is smaller than the size of the proton ($1/Q_0$). 
If the size of the pomeron becomes larger the full 
triple pomeron vertex \cite{bart,blw} 
has to be taken into account.

The main concern in this paper is the exact convolution of
the Lipatov eigenfunction with the meson wave function.
At large enough energies
the azimuthally symmetric solution ($n=0$) gives the dominant
contribution and we can neglect eigenfunctions with $n\ne 0$.
We study the resulting
$t$-dependence and look for a possible shrinkage in the $t$-shape
with increasing energy. Comparison will be made 
with the previous analytic calculations and numerical studies \cite{fr}.

\section{The exact BFKL-solution}
Let us begin by defining the cross section for the diffractive
scattering of a virtual photon off a proton:
\begin{eqnarray}
\frac{d \sigma(\gamma p \to V X)}{dtdx'}&=&\left( \frac{81}{16}G(x',t)+
\sum_f(q(x',t)+\bar{q}(x',t))\right)\;\frac{d \sigma(\gamma q \to V
q)}{dt} \nonumber\\
\frac{d \sigma(\gamma q \to V q)}{dt} &=& \frac{16 \pi}{81}
\frac{\alpha_s^4}{t^4} \left| {{\cal F}}(z,\tau) \right|^2.
\end{eqnarray}
Since the coupling of the BFKL-Pomeron to gluons is the same as 
the coupling to quarks except for a constant colour factor (81/16), we
have introduced, as our basic subprocess, the scattering off quarks only.
We have chosen to factorise the $1/t^4$ behaviour since it allows us to
write the remainder of the cross section in terms of the scaling variable
$$\tau \equiv -t/(Q^2 + m_V^2).$$
$Q^2$ is the photon virtuality (it may be zero) and $m_V$ is the vector 
meson mass. The rapidity gap between the produced
vector meson and the scattered quark is proportional to the
`energy' variable $z$:
$$ z = \frac{3 \alpha_s}{2 \pi} \ln \frac{\hat{s}}{-t+Q^2+m_V^2}, $$
where $\hat{s}$ is the Mandelstam variable
representing the $\gamma q$ centre-of-mass energy ($\hat{s}=x'W^2$ with
$W$ being the total hadronic energy).
From the work of Lipatov \cite{lip}, we have
\begin{equation}
{{\cal F}}(z,\tau) = \frac{t^2}{(2\pi)^3} \int_{-\infty}^{\infty} 
d\nu
\frac{\nu^2}{(\nu^2+1/4)^2} e^{\chi(\nu)z} I^V_{\nu}(q) I^{q*}_{\nu}(q)
\end{equation}
where
\begin{equation}
\chi(\nu) = -4(\gamma_E + {\cal R}e\;\psi(1/2+i\nu))
\end{equation}
and
\begin{equation}
I^V_{\nu}(q) = \int \frac{d^2 k}{(2\pi)^2} V(k,q) \int d^2\rho_1 d^2\rho_2
\left[ \frac{(\rho_1 - \rho_2)^2}{\rho_1^2 \rho_2^2} \right]^{1/2+i \nu}
e^{i k\cdot \rho_1 + i(q-k)\cdot \rho_2}.
\end{equation}
The momentum transfer $-t = q^2$, and for the pomeron-meson vertex we use the
non-relativistic form factor
\begin{equation}
V(k,q) = \frac{{\cal C}}{2} \left( \frac{1}{q_{\|}^2 + (k-q/2)^2} -
\frac{1}{q_{\|}^2+q^2/4} \right),
\end{equation}
where $$ {\cal C}^2 = \frac{3 \Gamma^V_{ee} m_V^3}{\alpha_{em}} $$ and
$4q_{\|}^2 = Q^2 + m_V^2$.

In this approximation, there is no helicity-flip amplitude and the rate of
production of (longitudinally polarised) mesons off longitudinal photons
is simply $Q^2/m_V^2$ times the transverse cross section.

The factor $I^{q}_{\nu}(q)$ is computed using the subtraction prescription of
Mueller and Tang \cite{mt}, i.e.
\begin{equation}
I^q_{\nu}(q) = \int \frac{d^2 k}{(2\pi)^2} \int d^2\rho_1 d^2\rho_2
\left[ -\left(\frac{1}{\rho_1^2}\right)^{1/2+i \nu}
       -\left(\frac{1}{\rho_2^2}\right)^{1/2+i \nu}  \right]
e^{i k\cdot \rho_1 + i(q-k)\cdot \rho_2}.
\end{equation}
It is straightforward to evaluate eq.(6):
\begin{equation}
I^q_{\nu}(q) = -\frac{4\pi}{q} \frac{\Gamma(1/2-i\nu)}{\Gamma(1/2+i\nu)}
(q^2)^{i\nu} 2^{-2 i \nu}.
\end{equation}

We now turn to the more complicated evaluation of eq.(4). The $k$-integral can
be performed exactly (it gives a McDonald function). Subsequently, it is
convenient to use the variables $R = (\rho_1+\rho_2)/2$ and
$\rho = \rho_1 - \rho_2$. After a Feynman parametrisation:
\begin{eqnarray}
I^V_{\nu}(q) &=& \frac{{\cal C}}{4\pi}\int d^2\rho d^2R \int_0^1 dx \;
\frac{ \rho^{1+2 i \nu} K_0(q_{\|}\rho) [x(1-x)]^{-1/2+i \nu} }
     { [x(R-\rho/2)^2 + (1-x)(R+\rho/2)^2]^{1+2 i \nu} } \nonumber \\
&\times& e^{iq\cdot \rho/2 + iq\cdot(R-\rho/2)}
     \frac{\Gamma(1+2 i \nu)}{\Gamma^2(1/2 + i \nu)}.
\end{eqnarray}
By changing variables to $R' = R - (x-1/2) \rho$ we are able to perform the
$R'$ integral and the $\rho$ angular integration. We get
\begin{eqnarray}
I^V_{\nu}(q) &=& \frac{{\cal C}\pi}{\Gamma^2(1/2+i \nu)}
\left(\frac{q}{2}\right)^{2 i \nu} \int_0^1 \frac{dx}{\surd[x(1-x)]}
\nonumber \\ &\times& \int_0^{\infty} d\rho \; \rho^2 \,
J_0(q\rho(1/2-x)) \; K_0(q_{\|}\rho) \; K_{2 i \nu}(q\rho \surd[x(1-x)]).
\end{eqnarray}
The trick now is to use the following representation of the $\delta$-function:
\begin{equation}
\delta(\rho^2 - \rho'^2) = \int_{1/2-i \infty}^{1/2+i \infty}
\frac{ds}{2\pi i} \rho^{-1-2s} \rho'^{2s-1},
\end{equation}
to give
\begin{eqnarray}
I^V_{\nu}(q) &=& {\cal C} \frac{2 \pi}{\Gamma^2(1/2+i \nu)}
\left(\frac{q}{2}\right)^{2 i \nu} \int \frac{dx}{\surd[x(1-x)]}
\int d\rho  \int \frac{ds}{2 \pi i} \nonumber \\ &\times&
\rho^{1-2s} J_0(q\rho(1/2-x)) \; K_{2 i \nu}(q\rho \surd[x(1-x)]) \nonumber \\
&\times& \int d\rho' \rho'^{2s} K_0(\rho' q_{\|}).
\end{eqnarray}
After performing the $\rho'$ integral, we are able to isolate the important
$q/q_{\|}$ factor. Putting $\xi = q\rho$ we have
\begin{eqnarray}
I^V_{\nu}(q) &=& {\cal C}\int \frac{ds}{2\pi i} \int \frac{dx}{\surd[x(1-x)]}
\frac{2^{2s}}{q^3} \left( \frac{q}{q_{\|}}\right)^{1+2s}
\frac{\pi \Gamma^2(1/2+s)}{\Gamma^2(1/2+i \nu)}
\left( \frac{q^2}{4}\right)^{i \nu} \nonumber \\ &\times&  \int d\xi \;
\xi^{1-2s} J_0(\xi(1/2-x)) \; K_{2 i \nu}(\xi \surd[x(1-x)]).
\end{eqnarray}
The $\xi$ integral can be done to give
\begin{eqnarray}
I^V_{\nu}(q) &=& {\cal C} \frac{\pi}{\Gamma^2(1/2+i\nu)}
\int \frac{ds}{2\pi i} \int \frac{dx}{[x(1-x)]^{3/2-s}} \frac{1}{q^3}
\left(\frac{q}{q_{\|}}\right)^{1+2s} \left( \frac{q^2}{4} \right)^{i \nu}
\nonumber \\ &\times& \Gamma^2(1/2+s) \Gamma(1-s-i \nu) \Gamma(1-s+i \nu)
\nonumber \\ &\times&
F(1-s-i\nu,1-s+i\nu;1;-\frac{(1-2x)^2}{4x(1-x)}).
\end{eqnarray}
Changing variables to $y = (1-2x)^2$ allows the $x$ integral to be performed,
giving
\begin{eqnarray}
I^V_{\nu}(q) &=& {\cal C} \frac{\pi}{q^3}
\frac{\Gamma(1/2)\Gamma(1/2-i\nu)}{\Gamma^2(1/2+i\nu)\Gamma(1-i\nu)} \left(
\frac{q^2}{4}\right)^{i\nu} \int \frac{ds}{2\pi i} 2^{2(1-s)} \left(
\frac{q}{q_{\|}}\right)^{1+2s}\nonumber \\ &\times&
\Gamma^2(1/2+s) \Gamma(1-s-i\nu)\Gamma(1-s+i\nu) \nonumber \\ &\times&
{}_3F_2(s-i\nu,1-s-i\nu,1/2;1-i\nu;1).
\end{eqnarray}
Using Watson's theorem, this can be simplified to the important result:
\begin{eqnarray}
I^V_{\nu}(q) &=& {\cal C} \frac{\pi^2}{q^3}
\frac{\Gamma(1/2-i\nu)}{\Gamma(1/2+i\nu)} \left( \frac{q^2}{4} \right)^{i\nu}
\int_{1/2-i \infty}^{1/2+i \infty} \frac{ds}{2\pi i}
\left( \frac{q}{q_{\|}}\right)^{1+2s} 2^{2(1-s)} \Gamma^2(1/2+s)
\nonumber \\ &\times&
\frac{\Gamma(1-s-i\nu)}{\Gamma(s/2+i\nu/2+1/2)\Gamma(s/2-i\nu/2+1/2)}
\nonumber \\ &\times&
\frac{\Gamma(1-s+i\nu)}{\Gamma(1-s/2+i\nu/2)\Gamma(1-s/2-i\nu/2)}.
\end{eqnarray}

Putting equations (7) and (15) into (2) we arrive at our final answer,
\begin{eqnarray}
{\cal F}(z,\tau) &=& 4{\cal C} \int_{-\infty}^{\infty} d\nu
\frac{\nu^2}{(\nu^2+1/4)^2} e^{\chi(\nu)z} \int_{1/2-i \infty}^{1/2+i \infty}
\frac{ds}{2\pi i} \tau^{1/2+s} \nonumber \\ &\times&
\frac{\Gamma^2(1/2+s)\Gamma(1-s-i\nu)}
{\Gamma(s/2+i\nu/2+1/2)\Gamma(s/2-i\nu/2+1/2)}
\nonumber \\ &\times&
\frac{\Gamma(1-s+i\nu)}{\Gamma(1-s/2+i\nu/2)\Gamma(1-s/2-i\nu/2)}.
\end{eqnarray}

In fig.1, the solid lines show the result of performing the $s$ and $\nu$
integrals exactly, i.e. they correspond to the exact solution to the
BFKL equation for $n=0$. The $\tau$ dependence at different $z$ values is
shown.

\section{The limits $\tau \gg 1$ and $\tau \ll 1$}
The $s$-plane integration in eq.(16) is quite straightforward to compute. It
leads to a power series in $\tau$ (or 1/$\tau$). For $\tau > 1$, we must close
the contour in the left half plane and for $\tau \gg 1$ it is the double pole at
$s=-1/2$ which provides the leading power behaviour. Similarly, for $\tau < 1$,
we must close in the right half plane and the leading behaviour is determined
by the two poles at $s=1\pm i\nu$.

For $\tau \gg 1$, we have
\begin{eqnarray}
{\cal  F}(z,\tau) &=& {\cal  C}
\int_0^{\infty} d\nu \frac{\nu^2}{(\nu^2+1/4)^2}
\frac{e^{\chi(\nu)z}}{{\cosh}\pi\nu}
\frac{32\pi}{\Gamma^2(1/4+i\nu/2)\Gamma^2(1/4-i\nu/2)} \nonumber \\ &\times&
\left\{ \ln \tau - 2 \gamma_E - 2 {\cal R}e\;\psi(3/2+i \nu) +
\frac{1}{\nu^2+1/4} \right\}
\end{eqnarray}
and for $\tau \ll 1$
\begin{eqnarray}
{\cal  F}(z,\tau) &=& {\cal  C}\int_0^{\infty} d\nu \frac{\nu^2}{(\nu^2+1/4)^2}
e^{\chi(\nu)z} \nonumber \\ &\times& 2 {\cal  R}e\;\left\{ \frac{8
\tau^{3/2+i\nu}}{\surd \pi} \frac{\Gamma^2(3/2+i\nu)\Gamma(-2i\nu)}
{\Gamma(1/2-i\nu)\Gamma(1+i\nu)}\right\}.
\end{eqnarray}
The $\nu$ integrals can easily be performed numerically.

Fig.1 illustrates how remarkably well these approximate forms reproduce the
full solution (solid lines). For $\tau \lapproxeq 0.3$, eq.(18) reproduces
the full solution over the whole $z$ range (dashed line). For large $\tau$
the agreement is excellent for large enough $z$ -- even for $\tau$ as low
as $0.5$ for $z = 0.8$ (dash-dot line)! The vanishing of the cross section
at $\tau \approx 10.1$ for $z=0.2$ reflects a large negative contribution
to the amplitude from large $\nu$ (where the integral is very slowly
convergent) and indicates the break down of the approximation.

From eq.(17) we conclude that, at large enough $\tau$, the dependence upon
$\tau$ and $z$ factorizes, i.e. the $t$-dependence
is energy independent in this case. The situation, however, changes for
$\tau<1$. Eq.(18) is dominated by the exponent:
\begin{eqnarray}
\exp \left\{\chi(\nu)\,z\;+\;(3/2\pm i\nu) \,\ln \tau \right\}
\end{eqnarray}
We can make a simple estimate using the saddle point method
with two different approximations of the $\chi$-function. In order
that this exponent be dominant, we require $\ln 1/\tau$ and $z$ to
be large enough. 

In the first case we assume $\ln 1/\tau$ to be not too large compared to
$z$ and expand $\chi (\nu)$ arround $\nu=0$: 
$\chi(\nu) \approx 8 \ln 2 - 28 \zeta(3) \nu^2$. The saddle point is then 
\begin{eqnarray}
\nu\;=\;\pm \;i\; \frac{\ln(1/\tau)}{56 \zeta(3)\,z}\;\;.
\end{eqnarray}
This limit refers to the BFKL solution at large 
$z$. The formal limit of $z \to \infty$ will be discussed in the next 
section. \lq Large enough $z$' means that $z$ is
much larger than $\ln (1/\tau)/[56 \zeta(3)]$ so that the saddle
point lies in the region ($\nu \approx 0$).
The saddle point integration then leads to
\begin{equation}
{\cal F}(z,\tau) \sim \frac{{\rm e}^{8 z \ln 2}}{z^{3/2}}
\exp \left( - \frac{\ln^2 \tau}{112 z \zeta(3)} \right).
\end{equation}

The second case we consider is for $\ln 1/\tau \gg z$. In this limit, we
need to expand the $\chi$ function around $\nu \approx \pm i/2$: $\chi(\nu) 
\approx 2/(1/2 \pm i\nu)$. The saddle point now reads:
\begin{eqnarray}
\nu = \pm \frac{i}{2} \left( 1 - \sqrt{\frac{8 z}{\ln 1/\tau}} \right).
\end{eqnarray}
The solution for the amplitude is then the well known double leading log 
result, i.e.
\begin{eqnarray}
{\cal F}(z,\tau) \sim \tau^2 \frac{\exp\{2 \sqrt{2z\ln (1/\tau)}\}}{\sqrt{z}}.
\end{eqnarray}
This result has an intuitive interpretation. For small $-t$ and large
$Q^2+m_V^2$ the dominant contribution comes from strongly ordered
transverse momenta within the BFKL-Pomeron, starting at $-t$ 
and ending at $Q^2+m_V^2$. As soon as the internal transverse momenta
are larger than $-t$ the effect of $t$ in the loop integrations is
negligible and the solution is identical to the $t=0$ evolution with $-t$ 
entering only as the starting scale. The cross section scales as expected, 
i.e. because of the $\tau^2$ factor it behaves like 
${\cal C}^2/(Q^2+m_V^2)^4$. Thus it is strongly suppressed at large $Q^2$ 
(it is a higher twist contribution). 

In order to estimate the magnitude of the shrinkage we calculate an effective
slope $\alpha'_{eff}$ following the soft Pomeron approach by differentiating 
the logarithm of the amplitude with respect to $t$ and $\ln \hat{s}$, i.e.
\begin{equation}
\alpha'_{eff} = -\frac{3 \alpha_s}{2 \pi} \frac{1}{Q^2+m_V^2}
\frac{\partial^2}{\partial z \partial \tau} \ln {\cal F}(z,\tau).
\end{equation}

When $\ln 1/\tau \ll 56 \zeta(3) z$ we can use Eq.(21). In which case,
\begin{equation}
\alpha'_{eff} = \frac{1}{|t|} \frac{3 \alpha_s}{2 \pi} \frac{\ln 1/\tau}{
56 z^2 \zeta(3) }.
\end{equation}

When $\ln 1/\tau \gg 8 z$ we can use Eq.(23). In which case,
\begin{equation}
\alpha_{eff}' =  \frac{1}{|t|} \frac{3 \alpha_s}{ 2\pi}
\frac{1}{\sqrt{2\;z \ln (1/\tau)}}.
\end{equation}

A different approach \cite{LevRys}, based on the 
diffusion in the impact parameter space leads to a dependence of 
$\alpha'_{eff}$ on $\alpha_s$ and $\ln s$ similar to Eq.(26).
(i.e. $\alpha'_{eff} \sim \sqrt{\alpha_s/\ln s}$).

In fig.2 we plot the energy dependence of the dimensionless quantity,
$-\partial^2 \ln {\cal F}/\partial z \partial \tau$.
The effective slope is obtained simply by multiplying by
$3 \alpha_s \tau /(2 \pi |t|) $. The solid line corresponds to the slope
extracted from Eq.(18). The dotted line corresponds to the effective slope
given by Eq.(25) and the dashed line is given by Eq.(26). 
From the figure, we can see clearly that the effective slope is small
for all accessable values of $\tau$ and $z$, i.e. $\alpha'_{eff} \lapproxeq 0.1$
GeV$^{-2}$ for $-t \gapproxeq 1$ GeV$^2$ and $z \gapproxeq 0.1$.
Recall that the slope of the soft pomeron trajectory advocated by
Donnachie \& Landshoff is $\approx 1/4$ GeV$^{-2}$ \cite{DL}.
Also we note that the saddle point evaluations do not lead to very good 
agreement with the full solution for sensible values of $z$ and $\tau$ (we 
have checked that the saddle point evaluations do eventually agree with the 
numerical evaluation for large enough $z$ (eq.(25)) or small enough $\tau$
(eq.(26))). 

To illustrate the shrinkage in a more familiar way, in fig.3 we
show the actual $t$-distribution for $J/\Psi$ photoproduction ($Q^2 = 0$).
The distribution is shown for the $\gamma q \to V q$ process at 
different $\gamma q$ energies, $\hat{s}$. For comparison the cross
sections at each energy are normalised to their value at $-t = 1$ GeV$^2$.
We took $\alpha_s = 1/4$ and $\hat{s} = W^2/3$ where $W = $ 50 GeV 
(solid line), 100 GeV (broken line) and 200 GeV (dotted line). These
energies are typical of the HERA kinematics. The corresponding $\gamma q$ 
cross sections at $-t = 1$ GeV$^2$ are: 16 nb/GeV$^2$, 72 nb/GeV$^2$ and 
360 nb/GeV$^2$ respectively. The $\gamma p$ cross sections are (in the 
valence quark region of large $x'\sim 1/3$) approximated by the individual 
quark cross sections multiplied by a factor of 3.  Note that the $z$
range probed by these energies is $0.5 \lapproxeq z \lapproxeq 0.9$. 

\section{The limit $z \gg 1$ at fixed $\tau$}
The asymptotic BFKL limit is that of $z \gg 1$ at fixed $\tau$. 
In the region where eq.(20) is valid we can present 
fully analytic formulae for large and small $\tau$, by recognising that the
dominant contribution to the $\nu$-integral only arises from the region 
around $\nu= 0$. As already detailed above, in this limit, 
$\chi(\nu) \approx 8 \ln 2 - 28 \zeta(3) \nu^2$ and so
\begin{equation}
\int_0^{\infty} d\nu \nu^2 e^{\chi(\nu)z} \approx \frac{\pi^{1/2}}{32}
\frac{e^{8 z \ln 2}}{[7 z \zeta(3)]^{3/2}}.
\end{equation}
Thus we obtain, for $\tau \gg 1$:
\begin{equation}
{\cal   F}(z,\tau) \approx {\cal  C}\frac{16\pi^{3/2}}{\Gamma^4(1/4)} 
\ln 16\tau \frac{{\rm e}^{8 z \ln 2}}{[7 z \zeta(3)]^{3/2}}.
\end{equation}
This is as in Eq.(21) after assuming $z$ to be large enough to
neglect any $z$-dependence of the $t$ behaviour.

For $\tau \ll 1$:
\begin{equation}
{\cal  F}(z,\tau) \approx {\cal   C} \;\sqrt{\pi}\; 
[ \ln 1/\tau  + 6\ln 2 - 4 ]
\tau^{3/2} \frac{e^{8 z \ln 2}}{[7 z \zeta(3)]^{3/2}}.
\end{equation}

Eqs. (28,29) were also found in ref.\cite{fr}. The $z \gg 1$ limit is
represented by the dotted line in fig.1, and can be seen to provide a 
faithful representation of the exact solution for $z \gapproxeq 0.8$. We 
should point out that the dotted lines in the figure actually correspond to 
the numerical evaluation of the amplitude in the $\nu = 0$ limit, 
i.e. they are valid for all $\tau$. However, outside of the region 
$0.2 \lapproxeq \tau \lapproxeq 1$ eqs.(28,29) produce essentially the same 
results.

\section{Conclusions}
We have shown that the exact analysis of the BFKL-pomeron 
in the azimuthally symmetric approximation ($n$=0) leads to a small 
shrinkage of the $t$-dependence at small $\tau$ ($\lapproxeq 0.1$) which
dies away with increasing $z$. For $z \gapproxeq 1$ the asymptotic
formulae of eqs.(28,29) start to become reasonable.

At small $z \lapproxeq 0.5$ and small $\tau \lapproxeq 0.5$, the exact 
$n=0$ solution is in nice agreement with the results of a fixed order 
BFKL summation, represented by the diamonds in fig.1. The diamonds were
first presented in ref.\cite{fr} and correspond to computing numerically
the BFKL corrections to the Born (2-gluon exchange) amplitude up to and 
including terms $\sim z^3$. For larger $z$, it was shown in ref.\cite{fr}
that the perturbation series is slowly convergent. This is why the
diamonds lie below the full $n=0$ solution. Certainly the numerical
results of ref.\cite{fr} strongly suggest that, for $\tau \lapproxeq 0.5$, the 
$n=0$ solution to the BFKL equation derived here is an excellent 
approximation to the full solution, even in the region of small $z$ 
(where the Born amplitude is sufficient). The region of small $\tau$
is experimentally accessable in light meson electroproduction and in heavy 
meson photo- and electro-production.

However, the situation changes when $\tau$ is increased. 
The diamonds in fig.1 no longer agree with the $n=0$ solution (solid line).
In ref.\cite{fr} it was found that the large $\tau$ region is
difficult to compute, and some progress was made in generating reliable
results. One of the main difficulties arises due to the presence of a
dip in the $\tau$-distribution \cite{fr}. The origin of this dip can be 
traced back to the peculiar behaviour of the Born amplitude, i.e. it vanishes 
for $\tau=1$. However, no dip is present in the $n=0$ solution.
When $z$ is increased the dip moves towards larger $\tau$ (but does not
seem to disappear). More work is needed to understand the 
behaviour of the amplitude in the region of the dip. Experimentally,
the region of the dip is accessable in the high-$t$ photoproduction of light
vector mesons.

It is of great interest to extend the present calculation towards
smaller $-t$ ($\le Q_0^2$). This kind of extension requires one
to go beyond the Mueller-Tang approach, pursuing the improvements made in
\cite{bart}. We expect the shrinkage in the $t$-dependence to become 
stronger \cite{blw}. 

The first HERA data on the the diffractive photoproduction of $J/\Psi$ mesons 
contains a number of events at high $p_T$ (i.e. $\gapproxeq 1$ GeV)
\cite{HERA}. The data sample is too small at present to draw any strong
conclusions and we look forward to the increase in statistics.

\section*{Acknowledgment}
We should like to thank Misha Ryskin for many important discussions.

\newpage
\section*{Figure Captions}
{\bf Figure 1:} The solid line represents the full BFKL-solution (eq.(16)), 
the dotted line shows the numerical evaluation of eq.(16)
about the $\nu \approx 0$ saddle point. The dashed line corresponds to
eq.(18) (i.e. $\tau \ll 1$) and the dash-dotted line to eq.(17) ($\tau \gg 1$).
The diamonds are from ref.\cite{fr} and correspond to a numerical
calculation of BFKL corrections to the Born contribution. \\ \ \\
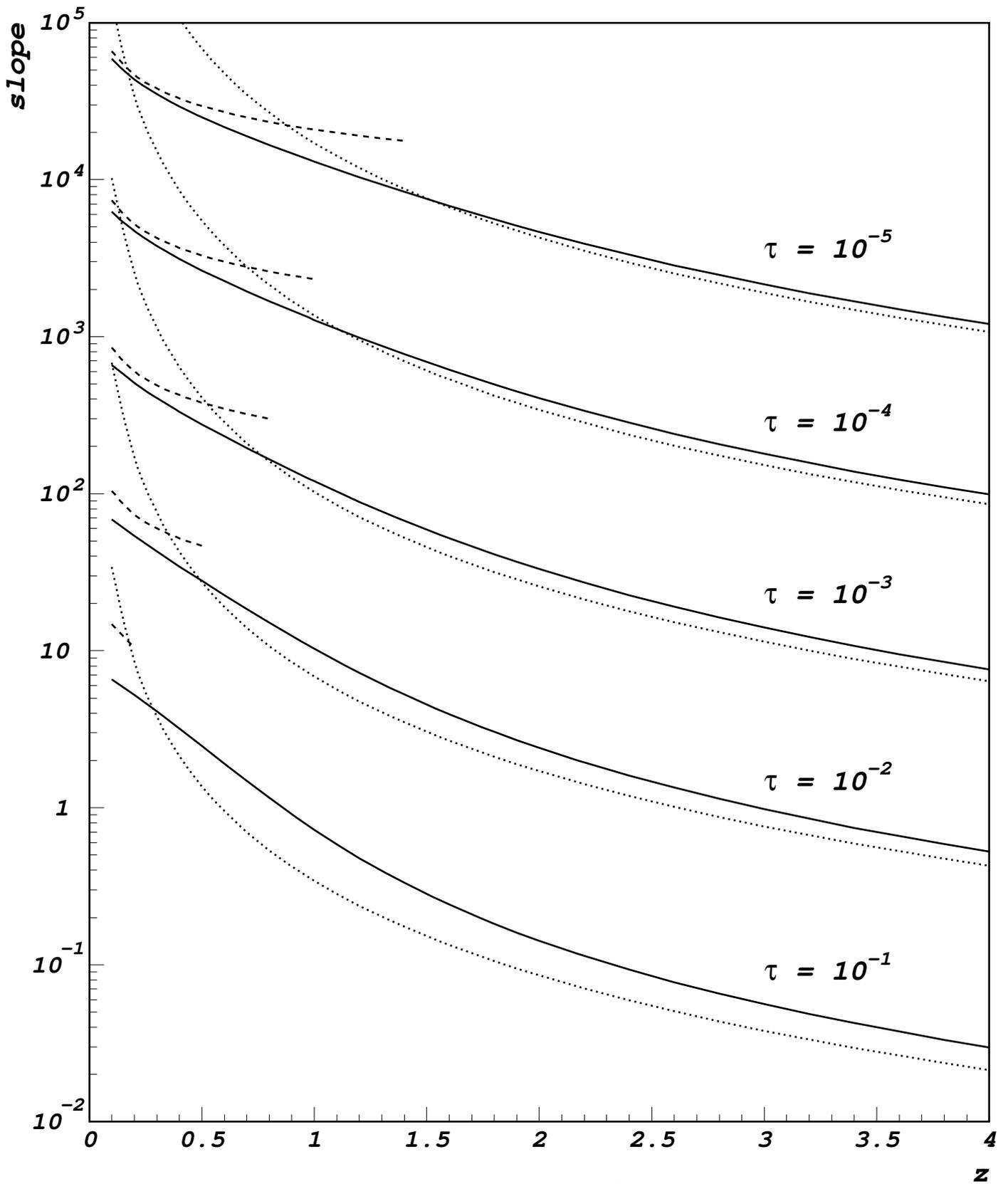
{\bf Figure 2:} The dimensionless $\alpha_{eff}'$ from eq.(24).
The solid line shows the numerical evalution based on eq.(18), the
dotted and dashed lines the corresponding saddle point solutions of
eqs.(25) and (26). \\ \ \\
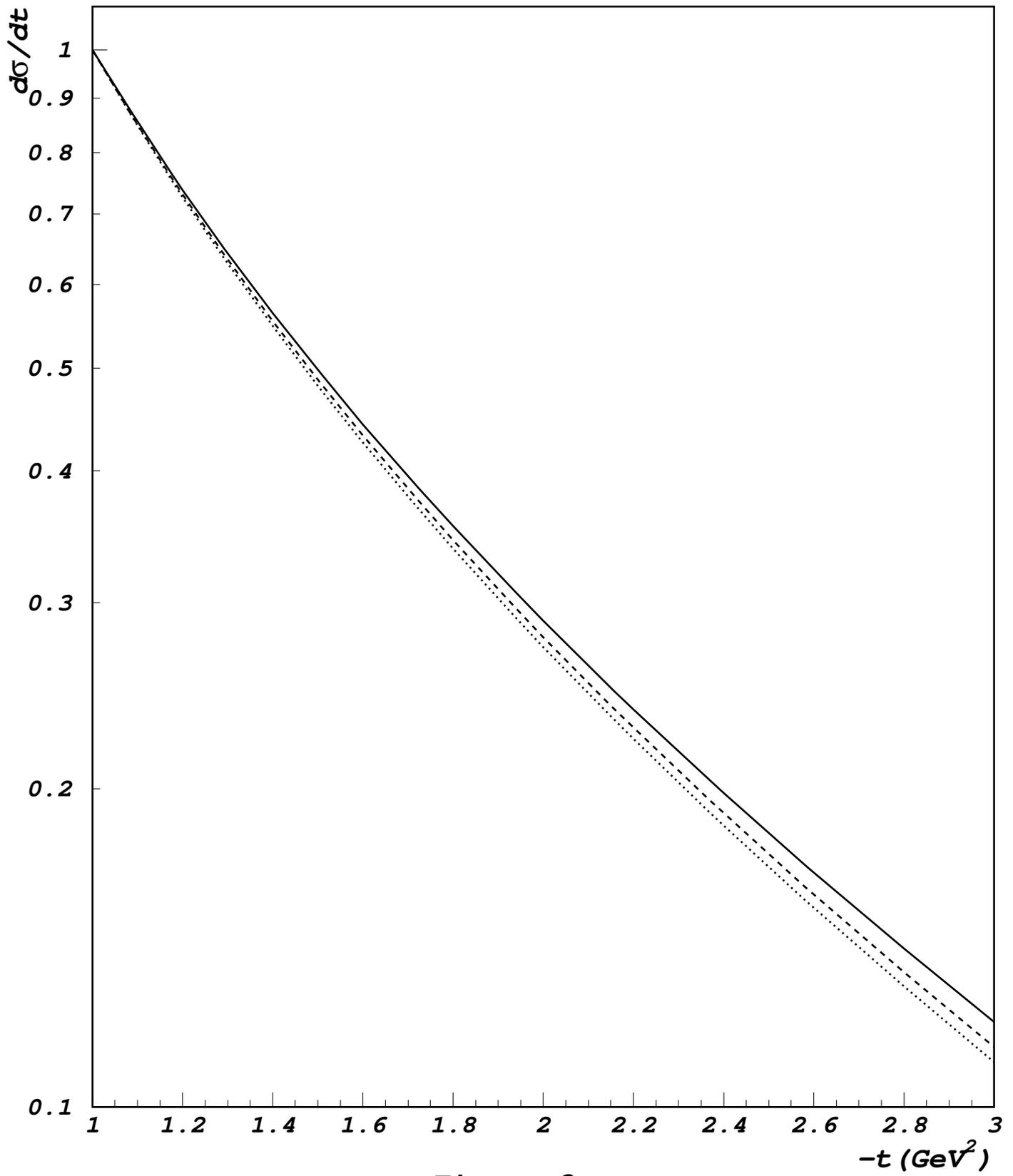
{\bf Figure 3:} The cross section normalized by its value at $|t|=1$ GeV$^2$ for
three different energies: $W$= 50 GeV (solid line), 100 GeV
(broken line) and 200 GeV (dotted line).


\begin{thebibliography}{99}

\bibitem{lip} L.N.Lipatov, Sov.Phys.JETP {\bf 63} (1986) 904.

\bibitem{fr} J.R.Forshaw and M.G.Ryskin, Zeit.Phys. {\bf C68} (1995) 137.

%

\bibitem{mt} A.H.Mueller and W-K.Tang, Phys.Lett. {\bf B284} (1992) 123.

\bibitem{bart} J.Bartels, J.R.Forshaw, H.Lotter, L.N.Lipatov,
M.G.Ryskin, M.W\"usthoff, Phys.Lett. {\bf B348} (1995) 589.

\bibitem{blw} J.Bartels, H.Lotter, M.W\"usthoff, Zeit.Phys. {\bf C 68}
(1995) 121.
\bibitem{LevRys} E.M.Levin, M.G.Ryskin, Zeit.Phys. {\bf C 48} (1990) 231.

\bibitem{DL} A.Donnachie and P.V.Landshoff, Phys.Lett. {\bf B123} (1983)
345; Nucl.Phys. {\bf B231} (1984) 189; Nucl.Phys. {\bf B267} (1986) 690.

\bibitem{HERA} H1 Collaboration: T.Ahmed et al, Phys.Lett. {\bf B338}
(1994) 507; ZEUS Collaboration: M.Derrick et al, Phys.Lett. {\bf B350}
(1995) 120. 

\end{thebibliography}
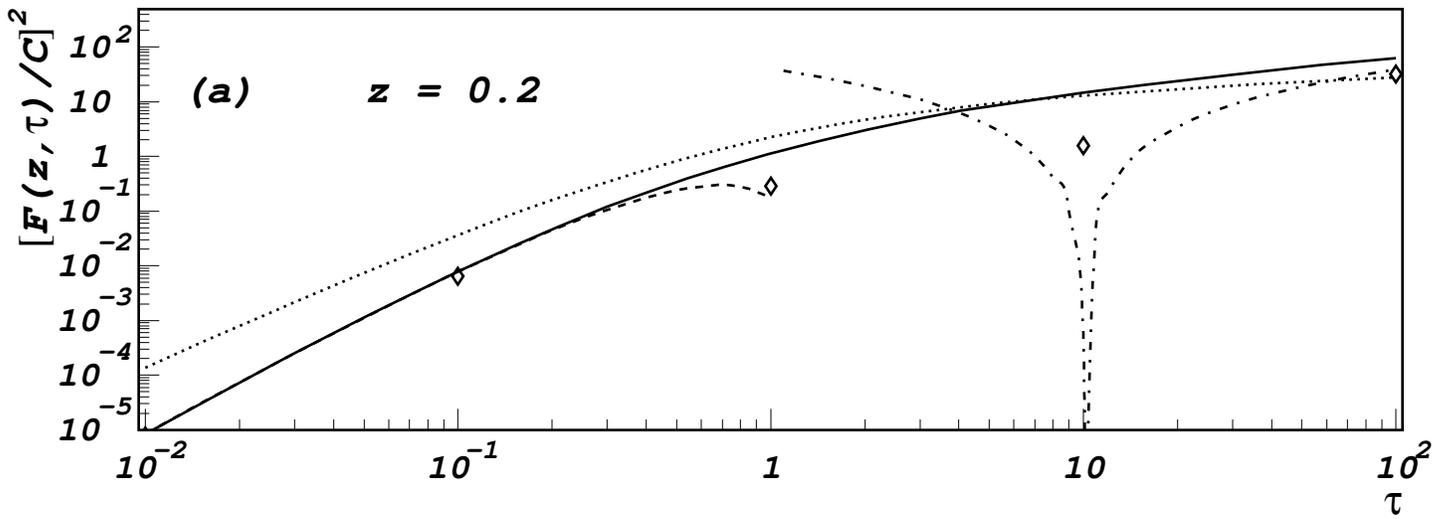
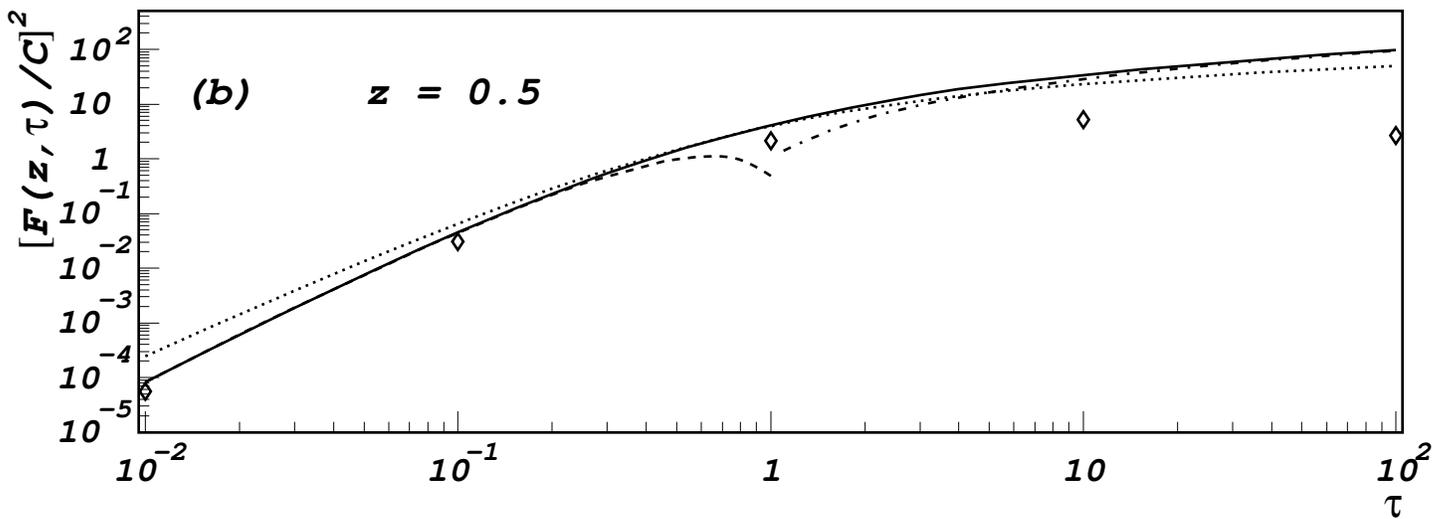
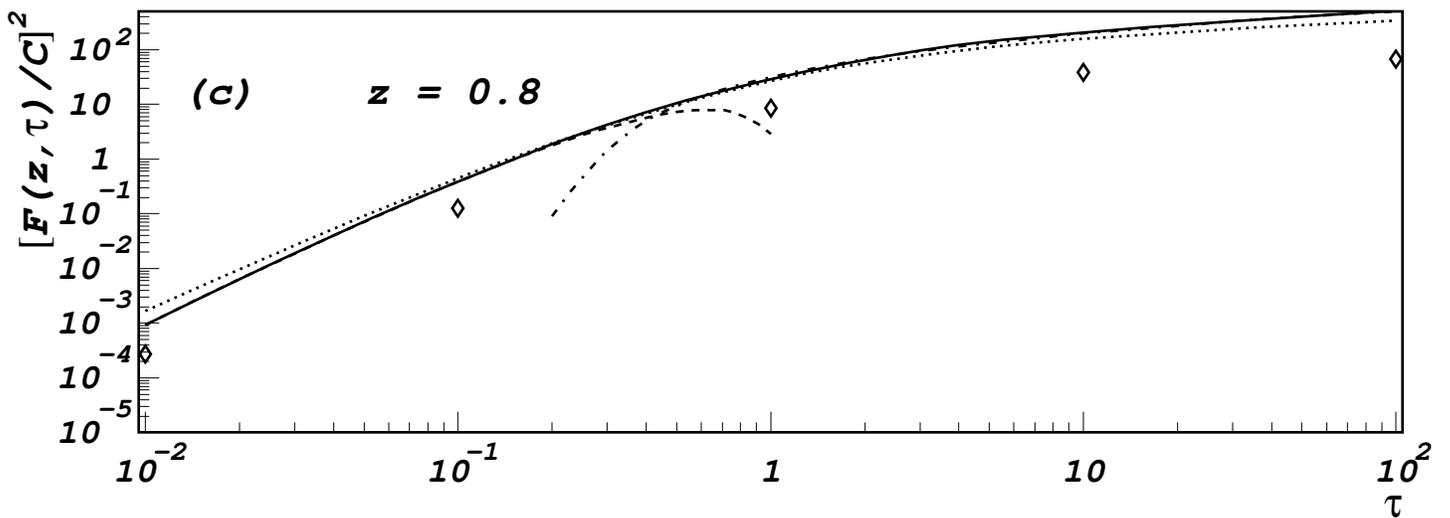
\end{document}